\documentclass[runningheads]{llncs}
\usepackage[utf8]{inputenc}
\usepackage[table,xcdraw]{xcolor}
\usepackage[T1]{fontenc}
\usepackage{enumitem}
\usepackage[colorlinks=true, linkcolor=blue!80, citecolor=blue!80, urlcolor=blue!80]{hyperref}
\usepackage{float}
\usepackage{flushend}
\usepackage[pdftex]{graphicx}
\usepackage{makecell}
\usepackage{multirow}
\usepackage{booktabs}
\usepackage{pifont}
\usepackage{marvosym} 
\usepackage{orcidlink}
\usepackage{varwidth}
\usepackage[most]{tcolorbox}
\usepackage{subcaption}

\newtcolorbox{source}[2][]{enhanced,
before skip=2mm,after skip=2mm,
colback=blue!5!white,colframe=black!50,boxrule=0.5mm,
colbacktitle=blue!85,
attach boxed title to top left={xshift=1cm,yshift*=1mm-\tcboxedtitleheight},
varwidth boxed title*=-3cm,
boxed title style={frame code={
\path[fill=tcbcolback!30!black]
([yshift=-1mm,xshift=-1mm]frame.north west)
arc[start angle=0,end angle=180,radius=1mm]
([yshift=-1mm,xshift=1mm]frame.north east)
arc[start angle=180,end angle=0,radius=1mm];
\path[left color=tcbcolback!60!black,right color=tcbcolback!60!black,
middle color=tcbcolback!80!black]
([xshift=-2mm]frame.north west) -- ([xshift=2mm]frame.north east)
[rounded corners=1mm]-- ([xshift=1mm,yshift=-1mm]frame.north east)
-- (frame.south east) -- (frame.south west)
-- ([xshift=-1mm,yshift=-1mm]frame.north west)
[sharp corners]-- cycle;
},interior engine=empty,
},
fonttitle=\bfseries,
title={#2},#1}

\begin{document}
\title{DFIR-Metric: A Benchmark Dataset for Evaluating Large Language Models in Digital Forensics and Incident Response}
\titlerunning{DFIR-Metric: A Benchmark Dataset for Evaluating LLMs in DFIR}

\author{
Bilel Cherif\inst{1}\orcidlink{0009-0006-0095-106X} \and
Tamas Bisztray\inst{2}\orcidlink{0000-0003-2626-3434} \and
Richard A. Dubniczky\inst{3}\orcidlink{0009-0003-3951-1932} \and
Aaesha Aldahmani\inst{1}\orcidlink{0009-0009-4075-5284} \and
Saeed Alshehhi\inst{1}\orcidlink{0009-0004-3061-2400} \and
Norbert Tihanyi\inst{1,3}\textsuperscript{(\Letter)}\orcidlink{0000-0002-9002-5935} 
}

\authorrunning{B. Cherif et al.}

\institute{
Technology Innovation Institute, Abu Dhabi, UAE\\
\email{\{bilel.cherif,saeed.alshehhi,aaesha.aldahmani,norbert.tihanyi\}@tii.ae}
\and
University of Oslo, Oslo, Norway\\
\email{tamasbi@ifi.uio.no}
\and
Eötvös Loránd University, Budapest, Hungary\\
\email{\{dubniczky,ntihanyi\}@inf.elte.hu}
}
\maketitle              
\begin{abstract}
Digital Forensics and Incident Response (DFIR) involves analyzing digital evidence to support legal investigations. Large Language Models (LLMs) offer new opportunities in DFIR tasks such as log analysis and memory forensics, but their susceptibility to errors and hallucinations raises concerns in high-stakes contexts. Despite growing interest, there is no comprehensive benchmark to evaluate LLMs across both theoretical and practical DFIR domains. To address this gap, we present DFIR-Metric, a benchmark with three components: (1) Knowledge Assessment: a set of 700 expert-reviewed multiple-choice questions sourced from industry-standard certifications and official documentation; (2) Realistic Forensic Challenges: 150 CTF-style tasks testing multi-step reasoning and evidence correlation; and (3) Practical Analysis:  500 disk and memory forensics cases from the NIST Computer Forensics Tool Testing Program (CFTT). We evaluated 14 LLMs using DFIR-Metric, analyzing both their accuracy and consistency across trials. We also introduce a new metric, the Task Understanding Score (TUS), designed to more effectively evaluate models in scenarios where they achieve near-zero accuracy. This benchmark offers a rigorous, reproducible foundation for advancing AI in digital forensics. All scripts, artifacts, and results are available on the project website at \url{https://github.com/DFIR-Metric}.


\end{abstract}

\section{Introduction}
Since the Turing Test first challenged machines to mimic human conversation~\cite{Turing2021-eq}, progress in \textit{Natural Language Processing (NLP)} has been tracked through various benchmarks. 
As noted by Wang et al.~\cite{wang-etal-2025-benchmark}, modern \textit{Large Language Models (LLMs)}, powered by neural networks and transformers~\cite{10.5555/3295222.3295349}, often record near-perfect scores on widely used suites such as GLUE and SQuAD~\cite{wang-etal-2018-glue,Rajpurkar2016-ly}, which reduces the effectiveness of these tests. In response, some new benchmarks like \textsc{Frontiermath}~\cite{glazer_frontiermath_2024} are made future-proof, and even advanced models can only achieve $1.7\%$ accuracy. These highly complex benchmarks do not support clear differentiation between the capabilities of current models.
LLMs hold immense potential for various fields, including cybersecurity~\cite{tihanyi_cybermetric_2024}, software engineering~\cite{10109345}, biomedicine~\cite{Chen2025-mh} or law~\cite{fei-etal-2024-lawbench}, which has sparked calls for privacy-aware, reliability-oriented, and domain-tailored benchmarks~\cite{wang-etal-2025-benchmark}.

\textit{Digital Forensics and Incident Response} (DFIR) is one such domain where practitioners analyze logs, e-mails, and multilingual reports to identify evidence, reconstruct timelines, and mitigate threats~\cite{Johansen2020-aw}. Recent studies show promising results when LLMs are applied in the DFIR domain, particularly for log filtering, artifact classification, and incident reporting~\cite{SHARMA2025301872,yin2025digitalforensicsagelarge,MICHELET2024301683,loumachi_advancing_2025,10654427}. However, the stakes are especially high. Errors can compromise evidence or misdirect investigations, and the use of proprietary models may violate strict confidentiality requirements. LLMs are known to hallucinate facts and misinterpret context~\cite{SOOD2025110307}. Before they can be trusted in DFIR workflows, we need rigorous, task-specific evaluations that measure not only one-off success through accuracy but also reliability and consistency.

Evaluating the performance of LLMs within the DFIR domain remains a significant challenge due to the absence of a comprehensive benchmark datasets and well-defined evaluation metrics. Although several strong general-purpose and domain-specific benchmarks are available, none provide a comprehensive evaluation across the diverse landscape of DFIR. As a result, practitioners lack a clear framework to determine when LLMs can be reliably applied and when expert validation is still required. A question naturally rises: ``\textit{Which specific DFIR tasks can LLMs effectively support,  and in which areas is human expertise still essential?}''
To obtain a detailed answer, we frame the study around the following research questions:

\begin{tcolorbox}[enhanced,title=Research Questions,
colframe=gray!50!black,colback=gray!10!white,
arc=1mm,
colframe=black!50,
fonttitle=\bfseries,coltitle=black!50!black,
attach boxed title to top text left=
{yshift=-0.50mm},
boxed title style={skin=enhancedfirst jigsaw,
size=small,arc=1mm,bottom=-1mm,
interior style={fill=none,
top color=gray!30!white,
bottom color=gray!20!white}}]
{\small
\begin{itemize}[leftmargin=*, label={}]
\item \textbf{RQ1:} What level of comprehension and confidence do LLMs exhibit in DFIR domain knowledge when challenged with certification-grade multiple-choice questions?
\item \textbf{RQ2:} To what extent can LLMs accurately and reliably solve practical forensic workflows, like log triage, memory-dump analysis, reverse engineering, and string search?
\item \textbf{RQ3:} Among the leading proprietary models and the strongest open-source alternatives, which achieve the highest scores in a unified evaluation?

\end{itemize}
}
\end{tcolorbox}
To the best of our knowledge, no comprehensive and standardized benchmark currently exists in the literature to thoroughly address these research questions. To fill this gap, we introduce \texttt{DFIR-Metric}, a novel suite of benchmark tasks and datasets to evaluate LLMs in the DFIR domain. According to NIST Special Publication 800-86 \textit{"Guide to Integrating Forensic Techniques into Incident Response"}~\cite{kent_guide_2006}, the digital forensics process consists of five key steps: identifying evidence, collecting artifacts, examining data, analyzing findings, and reporting results. Our benchmark evaluates LLMs on the first four stages, emphasizing technical accuracy and procedural rigor, while intentionally excluding the final legal reporting phase. This paper makes the following three key contributions:

\setcounter{footnote}{0}
\begin{itemize}
\small
    \item \textbf{DFIR-Metric}: We design a three-part dataset to evaluate LLMs on: (a) DFIR knowledge, using 700 human-verified multiple-choice questions sourced from industry certifications and official documentation; (b) practical disk and memory forensics tasks, evaluated through the string search tests of NIST’s \textit{Computer Forensics Tool Testing Program} (CFTT); and (c) CTF-style challenges on realistic forensic investigations that require planning, analytical reasoning, and evidence correlation.
    \item \textbf{Improved Evaluation Metrics}: Beyond single-pass accuracy, we assess each task multiple times to ensure reliability and introduce a new evaluation metric: the \textit{Task Understanding Score} (TUS), which rewards models for accurately completing steps in a multi-step pipeline;
    \item \textbf{Reproducability}: All artifacts, associated scripts, and the final \texttt{DFIR-Metric} dataset are available on the project's GitHub page, allowing independent researchers to integrate new models, replicate our results, and expand the evaluation as needed. We assessed 14 state-of-the-art LLMs to capture the current landscape of model advancements. (\url{https://github.com/DFIR-Metric}) 
\end{itemize}

The remainder of the paper is organized as follows. Section~\ref{sec:related_work} reviews related work. Section~\ref{sec:methodology} outlines the methodology used to construct the benchmark’s three components. Section~\ref{sec:experimentaL_result} presents experimental results for various state-of-the-art LLMs, while Section~\ref{sec:conclusion} concludes the paper.

\subsection*{Ethical Considerations}


All 700 DFIR-Metric questions were built from publicly available sources. Any text that resembled certification material was paraphrased or abstracted to prevent direct association with specific certification bodies or copyrighted material, and brief quotations are used only for research---a context generally covered by fair-use (or equivalent) provisions. The benchmark is independent of, and unendorsed by any certification body. Our aim is to support open, ethical research while respecting the rights of certification providers, content creators, and the broader digital forensics community.

\section{Related Work}
\label{sec:related_work}
Ferrag et. al~\cite{ferrag_generative_2025} identified nine main areas where LLMs are being used today marking DF as a standalone field. Sharma et al.~\cite{sharma_forensicllm_2025} proposed ForensicLLM, a model fine-tuned on a custom Q\&A dataset for digital forensics tasks, but neither the model nor its dataset is publicly released. Several recent studies have explored the use of pre-trained large language models for a variety of tasks across the digital forensic investigation pipeline. These tasks include timeline reconstruction~\cite{loumachi_advancing_2025}, automated report writing~\cite{michelet_chatgpt_2024}, and technical analyses such as malware detection and reverse engineering~\cite{ferrag_revolutionizing_2024,joyce_avscan2vec_2023}. Other work has focused on artifact examination~\cite{scanlon_chatgpt_2023}, as well as more practical applications like evidence extraction, scripting automation, and data recovery~\cite{scanlon_chatgpt_2023}. These studies highlight the growing interest in adapting LLMs to support various stages of forensic workflows, though many remain exploratory in nature. Wickramasekara et al.~\cite{wickramasekara_framework_2024} introduced AutoDFBench to assess AI coding skills in string search using NIST CFTT test suites. 
We note that Module III of \texttt{DFIR-Metric}, which will be detailed in the Section~\ref{sec:methodology}, also utilizes the NIST CFTT challenges to assess LLMs' capabilities in string search tasks. 

Most existing evaluation frameworks for LLM‐based forensics still rely on ad-hoc, chat-style prompt tests. This approach does not scale well to large question sets, provides limited control over response variability, and poses significant challenges for reproducibility. Horsman and Lyle~\cite{horsman_dataset_2021} emphasized the lack of high-quality datasets in digital forensics and proposed several guiding principles for dataset creation. Expanding on their work, we identify four key requirements that any robust forensic benchmark should meet: (i) publicly accessible, well-organized benchmarks hosted on platforms such as GitHub or Hugging Face; (ii) a plug-and-play evaluation framework that allows models to be tested via simple API integration; (iii) evaluation metrics that go beyond accuracy to include critical risks such as hallucination frequency and domain-specific blind spots; and (iv) a formal dataset specification that adopts a standardized format to support auditability and long-term reproducibility. While such datasets exist in broader cybersecurity domains, for example~\cite{tihanyi_dynamic_2024}, none of the existing DFIR-specific datasets fully meet all the criteria outlined above. Table~\ref{tab:dfir_datasets_cleaned} lists datasets, benchmarks, and frameworks that satisfy a subset of the criteria and support the DFIR field.
\definecolor{lightgray}{gray}{0.9}

\begin{table*}[ht]
\centering
\scriptsize
\caption{DFIR-Relevant Datasets and Benchmarks}
\label{tab:dfir_datasets_cleaned}
\renewcommand{\arraystretch}{1.2}
\rowcolors{2}{lightgray}{white}
\begin{tabular}{p{2.8cm} p{3.1cm} p{6cm}}
\toprule
\textbf{Dataset Name} & \textbf{Modality} & \textbf{Benchmark Scope / Description} \\
\midrule
\texttt{CyberMetric}~\cite{tihanyi_cybermetric_2024} & Textual (MCQ) & Cybersecurity benchmark with 10,000 MCQ;  \\
\texttt{DIA-Bench}~\cite{tihanyi_dynamic_2024} & Textual (JSON) & Cybersewcurity/math reasoning benchmark\\
\texttt{RAISE}~\cite{dang-nguyen_raise_2015} & Multimedia (Images) & Camera-native images to support classification tasks\\
\texttt{Vision Forensics}~\cite{shullani_vision_2017} & Multimedia (Video) & Device-attributed video samples for integrity analysis \\
\texttt{Timeline Analysis}~\cite{studiawan_towards_2025} & Timeline (CSV/JSON) & Plaso-based timeline QA benchmark for evaluating LLMs  \\
IoT-CAD~\cite{mohamed_iot-cad_2025} & Memory, Disk, Net & Labeled IoT attack traces with memory and network data. \\
\texttt{DeepSpeak}~\cite{barrington_deepspeak_2025} & Multimedia (Aud/Vid) & 100h webcam speech for deepfake detection tasks \\
\texttt{CIC-MalMem-2022}~\cite{carrier_detecting_2021} & Memory Dumps & 58k labeled Windows dumps (malware/benign) \\
\texttt{Unraveled}~\cite{myneni_unraveled_2023} & Logs (Net + Host) & Multi-week APT simulation with labeled detection logs. \\
\texttt{SCVIC-APT-2021}~\cite{liu_new_2022} & Network (pcap) & APT emulation with attack phases and labeled flows \\
\texttt{SCVIC-CIDS-2021}~\cite{liu_collaborative_2022} & Logs (Net + Host) & Host/network logs combining CIC-IDS-2018 traffic traces \\
\texttt{AutoDFBench}~\cite{wickramasekara_autodfbench_2025} & Disk + Artifacts & AI tool validation benchmark against NIST CFTT \\
\texttt{CTIBench}~\cite{alam_ctibench_2024} & Textual (TSV) & CTI benchmark with threat classification, CVSS scoring \\
\bottomrule
\end{tabular}
\end{table*}

\section{Methodology: Dataset Creation and Evaluation Metrics}
\label{sec:methodology}
\begin{figure}[h]
    \centering
    \begin{minipage}{\textwidth}
        \centering
        \includegraphics[width=\textwidth]{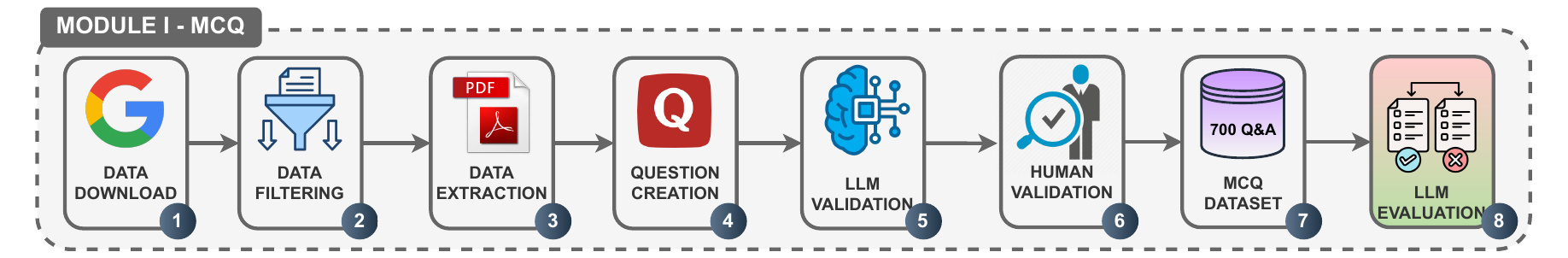}
        \subcaption{\textbf{Module I:} Multiple-Choice Question Dataset Generation and LLM Evaluation}\label{fig:module1}
        \vspace{1em}
        
        \includegraphics[width=\textwidth]{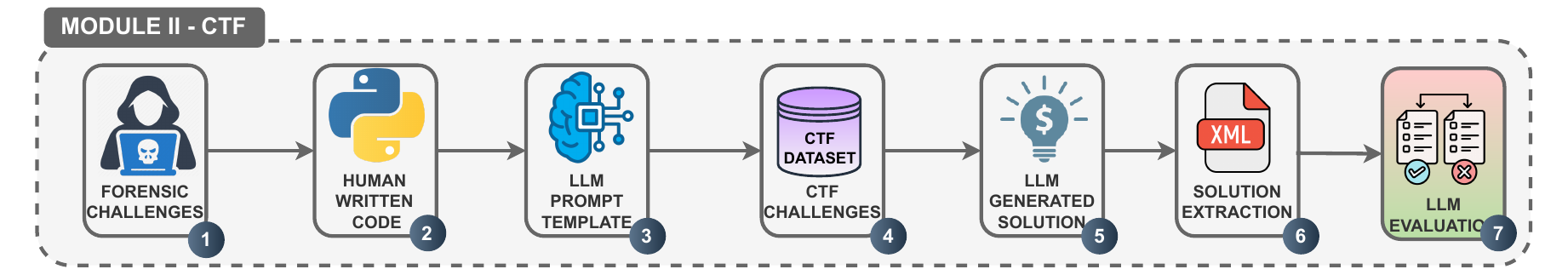}
        \subcaption{\textbf{Module II:} CTF-style Challenges Dataset Generation  and LLM Evaluation }\label{fig:module2}
        \vspace{1em}
        
        \includegraphics[width=\textwidth]{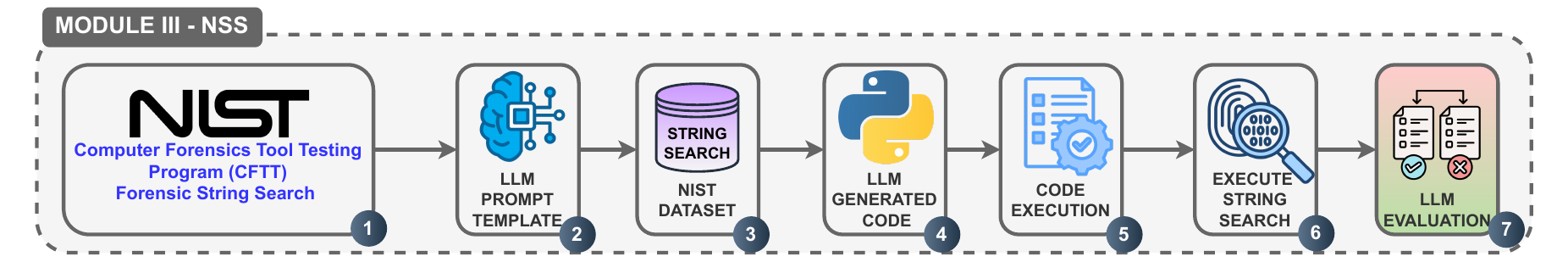}
        \subcaption{\textbf{Module III:} NIST String Search Dataset Generation  and LLM Evaluation }\label{fig:module3}
    \end{minipage}
    \caption{DFIR-Metric evaluation framework, consisting of three modules.}
    \label{fig:dfir-framework}
\end{figure}

\texttt{DFIR-Metric} consists of three core components, as shown in Figure~\ref{fig:dfir-framework}. Module I focuses on 700 multiple-choice question generation and evaluation. Module II contains CTF-style forensic challenges, covering a wide range of real-world scenarios from comprehensive log analysis to reverse engineering tasks. Module III introduces the NIST CFTT String Search Challenge\footnote{\scriptsize\url{https://www.nist.gov/itl/ssd/software-quality-group/computer-forensics-tool-testing-program-cftt}}, requiring LLMs to apply advanced forensic skills to analyze disk images and locate specific artifacts.

\subsection{Module I - Multiple-Choice Questions (Static)}

To assess theoretical competence in DFIR, we built a high-quality multiple-choice dataset aligned with international standards and certifications. An eight-step pipeline (Figure~\ref{fig:module1}.) harvested candidate questions from peer-reviewed articles, official guidelines, and certification exams, followed by an LLM grammar check and a 200-hour expert review. Ambiguous questions such as ```\textit{Where are deleted files stored in Windows operating systems?}'' were revised to eliminate imprecision. In Windows 10, deleted files reside in \texttt{C:\textbackslash\$Recycle.Bin}, whereas in Windows XP, they are located in \texttt{C:\textbackslash RECYCLER}. This module tests only knowledge through multiple-choice questions and does not assess the practical skills required to perform forensic tasks. Practical skillsets will be evaluated in the following modules. An example question is shown in the code snippet below:

\begin{tcolorbox}[colback=blue!5, colframe=black, boxrule=0.5mm, sharp corners]
       
       \texttt{Question example (Module 1):} Which command can provide the investigators with details of all the loaded modules on a Linux-based system?\\
       \textbf{A:} "plist mod -a",
      \textbf{B:} "lsof -m",
      \textbf{C:} "lsmod",
      \textbf{D:} "list modules -a".
    
        \end{tcolorbox}

\subsection{Module II - CTF-style Forensic Challenges (Dynamic)}

Inspired by \textit{Capture-the-Flag} (CTF) events, this module evaluates log analysis, cryptographic puzzles, and system-forensics skills. This is a dynamic module where each task is based on a hand-crafted template. Parameters such as log lines, keys, file system artifacts, and attacker actions can be randomized to generate multiple unique instances of the same task. In the evaluation we probe each task template three times to test the reliability of LLMs in solving specific tasks. Figure~\ref{fig:module2}. outlines the pipeline. All templates and solutions were manually audited, preserving real-world DFIR complexity while providing a controlled ground truth for rigorous, reasoning-centric assessment. Some of the CTF templates are modified versions from our previous work~\cite{tihanyi_dynamic_2024}, while several brand new task were added for forensics. An example question is shown  below:

\begin{tcolorbox}[colback=blue!5, colframe=black, boxrule=0.5mm, sharp corners]
         \texttt{Question example (Module 2):} Find the flag in this hex dump. Note: Characters are XOR'ed with 0x55 before hex encoding \textbf{0x0000: 3f d7 8c 31 78 e0 4d 00 4d 3b fb 69 71 66 9a 26 0x0010: 99 0f f3 a6 16 21 9b a5 82 36 5a 90 28} .....
        \end{tcolorbox}

\subsection{Module III - NIST Forensic String Search (Static)}

The third module introduces hands-on disk analysis tasks focused on string search, a fundamental forensic technique. This benchmark is based on the \textit{NIST Computer Forensics Tool Testing Program}'s technical documentation, originally designed to evaluate tools like EnCase and Magnet AXIOM using standardized datasets such as the String Search Test Data Set Package Version 1.1, which contains known content across various file systems. To adapt these challenges for LLMs, we reformulated each task into a prompt accompanied by a valid disk image, asking the model to generate a Python script to solve the given forensic problem. To assess performance, we developed an automated evaluation pipeline that analyzes disk images, extracts memory blocks, parses file systems, and recovers both active and deleted files. This output was used to construct ground truth baselines, which were rigorously validated by human experts. These baselines served as reference outputs for evaluating and comparing LLM-generated responses across tasks. The entire process is illustrated in Figure~\ref{fig:module3}.

\subsection{Defining Task Understanding Score for LLM evaluation}
In~\cite{tihanyi_dynamic_2024} four novel metrics were introduced; \textit{Reliability Score} (RS@k), \textit{Task Success Rate } (TSR@k), \textit{Confidence Index } (Conf@k), and \textit{Near Miss Score} (NMS@k). Let \( t \) represent a question template with variable parameters, and let \( \mathcal{T} = \{t_1, t_2, \dots, t_n\} \) denote the set of all such templates. Let $\mathcal{Q}(\mathcal{T}, k) = \{q_1, q_2, \dots, q_{n \times k}\}$ denote the set of unique questions, where each template from $\mathcal{T}$ is used to generate 
$k$ different questions and let \(\mathcal{S}_\mathcal{Q} = \{s_1, s_2, \dots, s_{n \times k}\}\) represent the set of solutions corresponding to \(\mathcal{Q}\). We have \(f : \mathcal{Q}(\mathcal{T}, k) \to \mathcal{S}_\mathcal{Q} \) such that \( f(q_i) = s_i \) for all \( i \in \{1, \dots, n\times k\} \).

\begin{definition}[\textbf{Reliability Score}]
The \emph{Reliability Score (RS@k)} over a dataset \(\mathcal{Q}(\mathcal{T}, k)\) is calculated as:
\begin{equation}\label{eq:reliability}
    \text{RS@}(k) = \frac{1}{k} \sum_{i=1}^{n \times k}\mathcal{A}_{i}
\end{equation}
where \(\mathcal{A}_{i}\) denotes the score assigned to answering \(q_i\), defined as \(\mathcal{A}_i = +1\) if \(s_i\) is correctly returned for \(q_i\), \(0\) if \(q_i\) is skipped, and \(-2\) otherwise.
\label{def:one}
\end{definition}

\begin{definition}[\textbf{Task Success Rate}] The \emph{Task Success Rate (TSR@$(t_i,k)$)} evaluates the number of correct answers for a given question template $t_i$ out of the $k$ generated instances, where the number of templates is $i \in \{1, 2, \dots, n\}$.
\begin{equation}\label{eq:confidence3}
    \text{TSR@}(t_i,k) = \sum_{j=1}^{k} \mathcal{B}_{j}
\end{equation}
where the value of \( \mathcal{B}_{j} \) is defined as $\mathcal{B}_{j} =+1$ if $s_j$ is returned for $q_j$, $0$ otherwise. 

\end{definition}

\begin{definition}[\textbf{Confidence Index}] The \emph{Confidence Index (Conf@k)} represents the percentage of question templates in a dataset where, for a given template \( t_i \), all \( k \) generated queries are successfully answered, 
\begin{equation}
    \text{Conf@}(k) = \frac{100}{n} \sum_{i=1}^{n} 
    \begin{cases} 
        1 & \text{if }  \text{TSR@}(t_i,k)=k 
        \\
        0 & \text{otherwise}.
    \end{cases}
\end{equation}
\end{definition}
\subsubsection{A new metric - The Task Understanding Score (TUS)}
Metrics such as TSR@k, Conf@k, and the traditional Pass@k assess whether a response to question $q_i$ is fully correct, but they do not account for cases where an LLM demonstrates partial success on a task. If LLMs score zero on a given task, it cannot establish a meaningful ranking, nor it will provide insight on how close models are to the correct solution. In reality, answers frequently contain some correct components, and we should also give credit for partial correctness. We want to move beyond simply classifying answers as correct or not, and introduce more granular scoring. Let \( \mathcal{C} = \{c_1, c_2, \dots, c_m\} \) be the set of key criteria, where \( m = |\mathcal{C}| \). Each criterion can represent various aspects---for instance, whether the Python code generated by an LLM executes correctly, or whether key steps essential to solving the problem are present. We can measure how many criteria are satisfied when solving question $q_i$. Let \( r_{ij} \in \{0, 1\} \) indicate whether the \( j \)-th criterion is satisfied in the solution to question \( q_i \). Then:

\begin{definition}[\textbf{Task Understanding Score}]
The \emph{Task Understanding Score (TUS@\(m\))} quantifies how well responses capture the essential components of a solution. It measures the average proportion of criteria satisfied across all evaluated responses.

\begin{equation}
    \text{TUS@}m = \frac{1}{|\mathcal{Q}|} \sum_{i=1}^{|\mathcal{Q}|} \left( \frac{1}{m} \sum_{j=1}^{m} r_{ij} \right)
\end{equation}
where  $|Q|$ is the total number of evaluated questions.
\end{definition}

Using TUS@m, we can evaluate the performance of LLMs on challenging tasks where traditional metrics like accuracy often yield a score of zero. Even in such cases, TUS@m enables us to capture partial correctness by assessing which predefined building block of a solution is satisfied in the response. For Module III tasks, the number of criteria $|\mathcal{C}|$ is set to four ($m=4$), with a dataset comprising $|\mathcal{Q}|=500$ NIST Forensic String Search challenge.

\section{Experimental results}
\label{sec:experimentaL_result}
To asses the capabilities of different LLMs on the newly introduced \texttt{DFIR-Metric} we conducted an experiment on multiple commercial and open-sourced models. 
\subsection{Module I - Multiple-Choice Questions}
We evaluated 14 state-of-the-art models on the MCQ dataset. Each question was asked 3 times, where correct answer was randomized between A, B, C or D to eliminate guessing. The best-performing model was GPT-4.1, closely followed by GPT-4o and Grok 3 with only marginal differences. Among the open-source, non-proprietary models, the best performer was Qwen-2.5 with 72 billion parameters. It achieved a Confidence Index (CI) of 84.29\% with  $k=3$
and a Mean Accuracy (MA) of 89.90\%, which is only 5\% lower than the state-of-the-art GPT models.  Table~\ref{tab:MCQ_results} displays the final results of the 14 LLMs tested.

\subsection{Module II - CTF-style Forensic Challenges}

Each CTF task was issued as a single prompt. Following Definition~\ref{def:one}, models earned $+1$ for a correct response, $0$ for skip, and $-2$ for an incorrect answer. All prompts, tasks, and the Google Colab code are published on our GitHub page to support easy and reproducible research. The evaluations were conducted via API, and no code execution was performed by the models in this module—consistent with their standard API capabilities. This contrasts with Module III, where Python code execution was preformed for the NIST string search tasks. Table~\ref{tab:CTF_results} presents the final results of the tested LLMs on the CTF challenges. 
\newpage
\begin{table}[H]
\small
\centering
\caption{LLM Performance on 700 MCQ dataset ($k{=}3$) (Sorted by CI)}
\label{tab:MCQ_results}

\begin{tabular}{lcccc|cc}
\toprule
\cmidrule(lr){6-7}
\textbf{Model} & \textbf{Company} & \textbf{Size} & \textbf{License} & \textbf{Open} & \textbf{CI} & \textbf{MA} \\
\midrule
\rowcolor{green!50}
GPT-4.1            & OpenAI        & N/A         & Proprietary & \textcolor{red}{\ding{56}}            & 89.34\% & 92.75\% \\
\rowcolor{green!45}
GPT-4o             & OpenAI        & N/A         & Proprietary & \textcolor{red}{\ding{56}}            & 88.92\% & 93.03\% \\
\rowcolor{green!43}
Grok 3             & xAI           & 2700B       & Proprietary & \textcolor{red}{\ding{56}}            & 88.22\% & 91.91\% \\
\rowcolor{green!35}
Claude 3.7 Sonnet  & Anthropic     & N/A         & Proprietary & \textcolor{red}{\ding{56}}            & 86.40\% & 91.58\% \\
\rowcolor{green!30}
Gemini 2.5 Flash   & Google        & N/A         & Proprietary & \textcolor{red}{\ding{56}}            & 85.41\% & 90.37\% \\
\rowcolor{green!25}
Qwen-2.5           & Qwen          & 72B         & Apache 2.0  & \textcolor{green!50!black}{\ding{52}} & 84.29\% & 89.80\% \\
\rowcolor{green!15}
DeepSeek V3        & DeepSeek AI   & 671B        & DeepSeek    & \textcolor{green!50!black}{\ding{52}} & 81.76\% & 89.25\% \\
\rowcolor{orange!10}
GPT-4o-mini        & OpenAI        & N/A         & Proprietary & \textcolor{red}{\ding{56}}            & 79.94\% & 85.78\% \\
\rowcolor{orange!15}
Llama 3.3          & Meta          & 70B         & Llama 3     & \textcolor{green!50!black}{\ding{52}} & 79.80\% & 86.49\% \\
\rowcolor{orange!20}
WizardLM 2         & Microsoft     & 8$\times$22B & Apache 2.0  & \textcolor{green!50!black}{\ding{52}} & 77.84\% & 84.53\% \\
\rowcolor{orange!25}
Gemma 3            & Google        & 27B         & Gemma       & \textcolor{green!50!black}{\ding{52}} & 77.13\% & 84.71\% \\
\rowcolor{orange!35}
Mixtral-8x7B       & Mistral AI    & 46.7B       & Apache 2.0  & \textcolor{green!50!black}{\ding{52}} & 71.11\% & 80.36\% \\
\rowcolor{orange!50}
Gemma 2            & Google        & 9B          & Gemma       & \textcolor{green!50!black}{\ding{52}} & 68.58\% & 79.85\% \\
\rowcolor{red!30}
Mistral-3B         & Mistral AI    & 3B          & Apache 2.0  & \textcolor{green!50!black}{\ding{52}} & 25.66\% & 55.86\% \\
\hline
\end{tabular}
\vspace{-40pt}
\end{table}

\begin{table}[H]
\small
\centering
\caption{DFIR-Metric CTF Performance ($k=3$) (Sorted by Confidence Index)}
\label{tab:CTF_results}

\begin{tabular}{lccc|ccccc}
\toprule
\textbf{Model} & \textbf{Company} & \textbf{Size} & \textbf{Open} & \textbf{Correct} & \textbf{Skipped} & \textbf{Wrong} & \textbf{RS} & \textbf{CI} \\
\midrule
\rowcolor{red!10} GPT-4.1        & OpenAI & N/A & \textcolor{red}{\ding{56}} & 47 & 0   & 103 & -53.0  & 28\% \\
\rowcolor{red!15} GPT-4o         & OpenAI & N/A & \textcolor{red}{\ding{56}} & 46 & 18  & 86  & -42.0  & 26\% \\
\rowcolor{red!20} DeepSeek V3    & DeepSeek AI & 671B & \textcolor{green!50!black}{\ding{52}} & 43 & 18  & 89  & -45.0  & 22\% \\
\rowcolor{red!25} Qwen-2.5       & Qwen & 72B & \textcolor{green!50!black}{\ding{52}} & 35 & 14  & 101 & -55.7  & 20\% \\
\rowcolor{red!30} Llama3.3       & Meta & 70B & \textcolor{green!50!black}{\ding{52}} & 33 & 16  & 101 & -56.3  & 20\% \\
\rowcolor{red!30} Grok 3         & xAI & 2700B & \textcolor{red}{\ding{56}} & 40 & 5   & 105 & -56.7  & 20\% \\
\rowcolor{red!35} Gemini 2.5-flash & Google & N/A & \textcolor{red}{\ding{56}} & 38 & 1   & 111 & -61.3  & 20\% \\
\rowcolor{red!35} GPT-4o-mini    & OpenAI & N/A & \textcolor{red}{\ding{56}} & 27 & 14  & 109 & -63.7  & 20\% \\
\rowcolor{red!40} Claude 3.7 Sonnet & Anthropic & N/A & \textcolor{red}{\ding{56}} & 18 & 0   & 132 & -82.0  & 12\% \\
\rowcolor{red!45} Mixtral-8x7B   & Mistral AI & 47B & \textcolor{green!50!black}{\ding{52}} & 24 & 1   & 125 & -75.3  & 12\% \\
\rowcolor{red!50} Gemma 3        & Google & 27B & \textcolor{green!50!black}{\ding{52}} & 22 & 2   & 126 & -76.7  & 10\% \\
\rowcolor{red!55} Mistral        & Mistral AI & 3B & \textcolor{green!50!black}{\ding{52}} & 22 & 2   & 126 & -76.7  & 10\% \\
\rowcolor{red!65} Gemma 2        & Google & 9B & \textcolor{green!50!black}{\ding{52}} & 0  & 0   & 150 & -100.0 & 0\% \\
\hline
\end{tabular}
\vspace{-40pt}
\end{table}

\begin{table}[H]
\small
\centering
\caption{NIST Forensic String Search ($m=4$) (Sorted by TUS@4)}
\label{tab:NSS_results}

\begin{tabular}{lccc|ccccc}
\toprule
\textbf{Model} & \textbf{Company} & \textbf{Size} & \textbf{Open} & \textbf{Correct} & \textbf{Syntax} & \textbf{Wrong} & \textbf{T/O} & \textbf{TUS@4} \\
\midrule
\rowcolor{red!10} GPT-4.1            & OpenAI      & N/A   & \textcolor{red}{\ding{56}} & 1 & 217 & 292 & 0   & 38.52\% \\
\rowcolor{red!25} GPT-4o             & OpenAI      & N/A   & \textcolor{red}{\ding{56}} & 0 & 283 & 226 & 1   & 27.99\% \\
\rowcolor{red!30} Gemini 2.5-flash   & Google      & N/A   & \textcolor{red}{\ding{56}} & 1 & 16  & 493 & 0   & 25.88\% \\
\rowcolor{red!35} Claude 3.7 Sonnet  & Anthropic   & N/A   & \textcolor{red}{\ding{56}} & 5 & 309 & 195 & 1   & 24.75\% \\
\rowcolor{red!40} DeepSeek V3        & DeepSeek AI & 671B  & \textcolor{green!50!black}{\ding{52}} & 0 & 268 & 13  & 229 & 22.40\% \\
\rowcolor{red!45} Grok 3             & xAI         & 2700B & \textcolor{red}{\ding{56}} & 0 & 316 & 194 & 0   & 21.71\% \\
\rowcolor{red!55} Llama3.3           & Meta        & 70B   & \textcolor{green!50!black}{\ding{52}} & 0 & 293 & 7   & 210 & 15.40\% \\
\rowcolor{red!60} GPT-4o-mini        & OpenAI      & N/A   & \textcolor{red}{\ding{56}} & 0 & 51  & 459 & 0   & 12.64\% \\
\rowcolor{red!65} Qwen-2.5           & Qwen        & 72B   & \textcolor{green!50!black}{\ding{52}} & 0 & 453  & 56  & 1 & 3.62\% \\
\hline
\end{tabular}

\end{table}

GPT-4.1 achieved the highest \emph{confidence index} (CI, 28\,\%), but its reliability score (RS) was lower than that of GPT-4o and \textsc{DeepSeek}~V3 because it attempted every task and accumulated 103 wrong answers, which highlights a potential architectural difference or system prompt design. The other GPT-4 variant skipped 18 uncertain items, incurring fewer penalties and posting the best reliability score ($-42$).

Among the non-proprietary, openly released models, \textsc{DeepSeek}~V3, Qwen-2.5, and Llama-3.3 form a tight cluster (CI 22–20\,\%), trailing GPT-4.1 by only 6–8 percentage points. Note, that the benchmark includes high-difficulty items---reverse engineering, cryptographic analysis, and large-scale log triage---that typically require code execution or multi-step reasoning. Although larger or more sophisticated models may one day solve these problems in a single pass, present systems should at minimum recognize their limits and abstain, rather than emit fluent yet incorrect answers. The frequency of such errors drives every model’s reliability score (range $-300$ to $150$) into negative territory. In this evaluation, partial success was not graded, however, in some cases models were able to extract the correct answer, but failed to return it in the requested format.

\subsection{Module III - NIST Forensic String Search}

We assessed the nine highest-scoring models from the MCQ benchmark on the NIST Forensic String Search task. The following task serves as an example; ``\textit{You are an experienced Digital Forensics and Incident response specialist. Answer the following question by writing a python script....}. Additionally, each prompt contained specific instruction like directing the LLMs to produce a self-contained Python script, and return its findings in a rigid JSON-like list of \texttt{<inode>:<filename>} pairs, prefixed by \texttt{DELETED} or \texttt{LIVE}. 

This setting stresses both technical competence (correct identification of strings in a forensic image) and compliance with a brittle output specification---two dimensions that everyday DFIR workflows routinely demand. To calcualte TUS@4, partial points can be awarded for the following categories: 
\textbf{(1)} determining the right offset of the file system from the prompt description, as each image contains 3 different file systems.
\textbf{(2)} properly identify the image path in the directory.
\textbf{(3)} identify the correct search string target, and if it requires regex or regular search.
\textbf{(4)} identifying the right extension for the artefact; doxc, txt, html, etc. Table~\ref{tab:NSS_results}  provides insights on how many occasions models were able to solve the task, or if they failed what was the main reason. The categories are: \textbf{Correct:} the script successfully extracts the target information from the forensic. \textbf{Wrong:} the script runs but fails to extract the correct datadisk image. \textbf{Timeout Execution:} the script does not complete within a predefined execution time. \textbf{Syntax Error:} the script fails to run due to code syntax issues.

Although GPT-4.1 secures the highest TUS@4 (38.5 \%), its advantage stems largely from a higher rate of \emph{partially} correct steps, rather than from wholesale task completion. Manual review revealed three recurrent error patterns across models: they sometimes hallucinate files, bash commands, paths or libraries that are absent from the image, causing the script to crash; even when the search logic is sound, the script may capture the wrong sub-string or omit a required field, producing only partially valid lines, and finally; tiny deviations from the rigid output schema, misplaced brackets, missing prefixes or commas invalidate otherwise correct answers.

Table \ref{tab:NSS_results} reports the results for the evaluation. TUS rewards incremental progress (correct code fragments, partially valid lists, etc.) rather than binary success. From a practitioner’s standpoint this nuance matters: a higher TUS model may still require substantial fine-tuning to yield admissible, reproducible evidence, but it has a step in the right direction.
Finally, we note that open-weight models (e.g.\ DeepSeek V3, Qwen-2.5, Llama 3.3) have yet to match the proprietary leaders in this task.

\section{Conclusion}
\label{sec:conclusion}
In this work, we introduce \texttt{DFIR-Metric}, the first extensive benchmark tailored to evaluate both theoretical knowledge and practical proficiency of LLMs in the domain of Digital Forensics \& Incident Response (DFIR). Spanning 700 certification-grade multiple-choice questions, a NIST-compliant string-search suite, and dynamically generated CTF investigations, \texttt{DFIR-Metric} evaluates models across the first four phases of the NIST 800-86 forensic workflow. To measure not only accuracy but also consistency and self-assessment, we used reliability metrics such as Confidence Index and Reliability Score for the evaluation, and introduced a novel metric---the Task Understanding Score (TUS)--- and executed every task multiple times for awarding partial task completion. Our work addressed three research questions:

\begin{itemize} \small
\item \textbf{RQ1}: \textit{What level of comprehension and confidence do LLMs exhibit in DFIR domain knowledge when challenged with certification-grade multiple-choice questions?}\\
\textbf{Answer}: The leading models demonstrate substantial mastery of core DFIR principles. GPT-4.1 achieves a Confidence Index of 89.34\% and a Mean Accuracy of 92.75\%. This underlines that high accuracy does not correspond with reliable problem solving, as models may guess and provide a correct answer by chance. This highlights the importance of repetitive testing and reliability metrics. The open-source Qwen-2.5-72B trails by only 5\%, indicating a narrowing proprietary edge, whereas compact models (e.g., Mistral-3B) perform scarcely above pure chance.
\item \textbf{RQ2}: \textit{To what extent can LLMs accurately and reliably solve practical forensic workflows such as log triage, memory-dump analysis, reverse engineering, and string search?}\\
\textbf{Answer}: Practical competence lags behind domain knowledge. In the NIST String Search module, no model produced meaningful results across the $500$ prompts, and even the top performer (GPT-4.1) achieved just 38\% partial-credit on the Task Understating Score, indicating incomplete pipeline execution (e.g., script generation succeeded but filesystem carving failed). In our CTF-style trials, the best model was again GPT-4.1, but solved only 28\% of tasks consistently. Notably, unlike other top performing models like GPT-4o, DeepSeek V3, or Qwen-2.5, GPT-4.1 was not able to skip any questions, highlighting severe limitations in comprehension and self reflection.

\item \textbf{RQ3}: \textit{Among the leading proprietary models and the strongest open-source alternatives, which achieve the highest scores in a unified evaluation?}\\
\textbf{Answer}: Overall, the proprietary models, GPT-4.1 and GPT-4o consistently lead in every Module: domain knowledge, CTF challenges, and NIST sting search tasks (although in the latter they were not able to solve a single task, and only achieve partial success through the task understanding score). Among the open source models Qwen–2.5 and DeepSeek V3 perform best in the multiple choice questions, Llama 3.3, WizardLM 2 and Gemma 3 are not trailing far behind. Interestingly, in the CTF challenges DeepSeek V3 performs very close to GPT-4o, skipping the same amount of questions and only getting a 4\% worse Confidence Index.

\end{itemize}
Our findings highlight steady progress but also underscore unresolved challenges in automating end-to-end DFIR investigations. Current LLMs can recall certification material and generate competent forensic scripts, yet struggle with sustained deductive reasoning, rigorous chain-of-custody logic, and calibrated confidence. Here it is important to highlight that we did not include reasoning models in the evaluation like o4-mini or DeepSeek R1, where we expect these models to perform slightly better based on~\cite{tihanyi_dynamic_2024}.

DFIR-Metric fills a critical evaluation gap, offering the community an open, extensible framework to measure future advances. We release all datasets, grading code, and baseline results to foster reproducibility and encourage iterative enhancement. We conclude that practical digital forensic scenarios and end-to-end forensic workflows remain out of reach for current models.

\begin{credits}
\subsubsection{\ackname} 
This research is supported and funded by the Technology Innovation Institute (TII), Abu Dhabi. Additional support is provided by ZEISS Digital Innovation; TKP2021-NVA Funding Scheme under Project TKP2021-NVA-29; ELTE-OTP Cyberlab—a collaboration between Eötvös Loránd University (ELTE) and OTP Bank Plc; EPSRC grant EP/T026995/1 titled “EnnCore: End-to-End Conceptual Guarding of Neural Architectures” under the Security for All in an AI-enabled Society program; the Research Council of Norway Project No. 312122 “Raksha: 5G Security for Critical Communications”; funding from Horizon Europe under Grant Agreement No. 101120853; and funding under Grant Agreement No. 101145874, supported by the European Cybersecurity Competence Centre.

\subsubsection{\discintname}
\textbf{Competing Interests:} The authors declare no competing interests relevant to the content of this article.
\end{credits}
%
%
%

\bibliographystyle{splncs04}
\bibliography{main}

\end{document}